# Identifying the Attractors of Gene Regulatory Networks from Expression Data under Uncertainty: An Interpretable Approach

Alireza Rowhanimanesh

*Faculty of Novel Interdisciplinary Technologies, University of Neyshabur*
*E-mail: rowhanimanesh@neyshabur.ac.ir*

**Abstract.** In systems biology, attractor landscape analysis of gene regulatory networks is recognized as a powerful computational tool for studying various cellular states from proliferation and differentiation to senescence and apoptosis. Therefore, accurate identification of attractors plays a critical role in determination of the cell fates. On the other hand, in a real biological circuit, genetic/epigenetic alterations as well as varying environmental factors drastically take effect on the location, characteristics, and even the number of attractors. The central question is: Given a temporal gene expression profile of a real gene regulatory network, how can the attractors be robustly identified in the presence of huge amount of uncertainty? This paper addresses this question using a novel approach based on Zadeh's *Computing with Words*. The proposed scheme could effectively identify the attractors from temporal gene expression data in terms of both fuzzy logic-based and linguistic descriptions which are simply interpretable by human experts. Therefore, this method can be considered as an effective step towards interpretable artificial intelligence. Without loss of generality, genetic toggle switch is considered as the case study. The nonlinear dynamics of this benchmark gene regulatory network is computationally modeled by the notion of uncertain stochastic differential equations. The results of in-silico study demonstrate the efficiency and robustness of the proposed method.

*1. Introduction*

In the valuable papers of Prof. Zadeh, the father of fuzzy logic, fuzzy computing has been proposed as a powerful tool for perception-based computing [1-4]. Although in the era of Industry 4.0, the emergence of novel interdisciplinary technologies such as artificial intelligence (AI) and Internet of Things (IoT) significantly affect medical sciences, medicine undoubtedly needs explainable AI which can be interpretable for human experts. Indeed, fuzzy logic is a crucial approach towards developing interpretable AI, from conventional clinical applications [5-6] to future nanomedical applications [7-12]. On the other hand, the role of intelligent analysis of omics data in the realization of precision medicine is very promising, and the recent advances in systems biology clearly reveals that how AI can help in whole-cell modeling, cell fate control, and cellular reprogramming [13-21].

In the context of gene regulatory networks (GRNs), one of the most applied biological circuits is Genetic Toggle Switch (GTS) that can be implemented inside a biological cell using biotechnology [22]. As displayed in Fig.1, in the absence of inducers, two stable states are possible: a) repressor 2 is transcripted by promoter 1, and b) promoter 2 transcribes repressor 1. According to the inhibitory interactions between these two repressors and the existence of noise and uncertainty in the circuit structure and the environment, bi-stability is the key property of this GRN. Since GTS is able to perform logical functions at molecular level and can be considered as a building block for construction of more complex GRNs, this basic circuit has been recognized as a benchmark GRN in the literature. Various researches has been performed on computational modeling and analysis of the dynamics of GTS and it has been demonstrated that it is nonlinear, stochastic, and very sensitive to uncertainty [23-25].

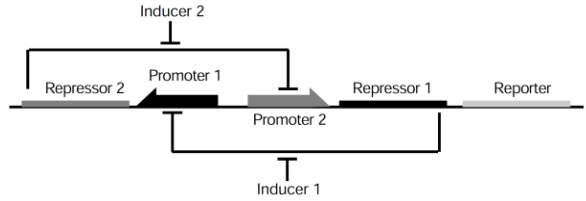

Fig. 1. The Gene Regulatory Network of a Genetic Toggle Switch (taken from [22])

## 2. A Robust Model in the Presence of Uncertainty

The nonlinear dynamics of the genetic toggle switch in Fig.1 can be deterministically modeled by a set of ordinary differential equations (ODEs) as follows [22]:

$$\begin{cases} \frac{dx_1}{dt} = \frac{\alpha}{1+x_2^\beta} - \rho x_1 \\ \frac{dx_2}{dt} = \frac{\alpha}{1+x_1^\beta} - \rho x_2 \end{cases} \quad (1)$$

where $x_i$ is the expression of repressor $i$ ($i = 1.2$), $\alpha$ is the effective rate of synthesis of repressor, $\beta$ is the cooperativity of repression of promoter, and $\rho$ is the degradation coefficient. Assume that $\alpha = 5$, $\beta = 2$, and $\rho = 1$. Then, the model has three equilibrium points including $(0.21, 4.79)$, $(1.52, 1.52)$, and $(4.79, 0.21)$. As depicted in Fig.2-a, the eigenvalues of the linearized model around the second equilibrium point are $\lambda_1 = 0.39$, and $\lambda_2 = -2.39$, which is corresponding to a saddle point behavior around this unstable equilibrium point in the phase portrait. The eigenvalues corresponding to the other equilibrium points are $\lambda_1 = -0.6$, and $\lambda_2 = -1.4$, that shows the existence of two stable nodes and the bistability property in the model. It is important to note that the deterministic model of Eq.1 does not handle the real-world uncertainty. According to this model and as displayed in Fig.2-b, there is a stable manifold around the saddle point, while it is not practical due to noise and uncertainty.

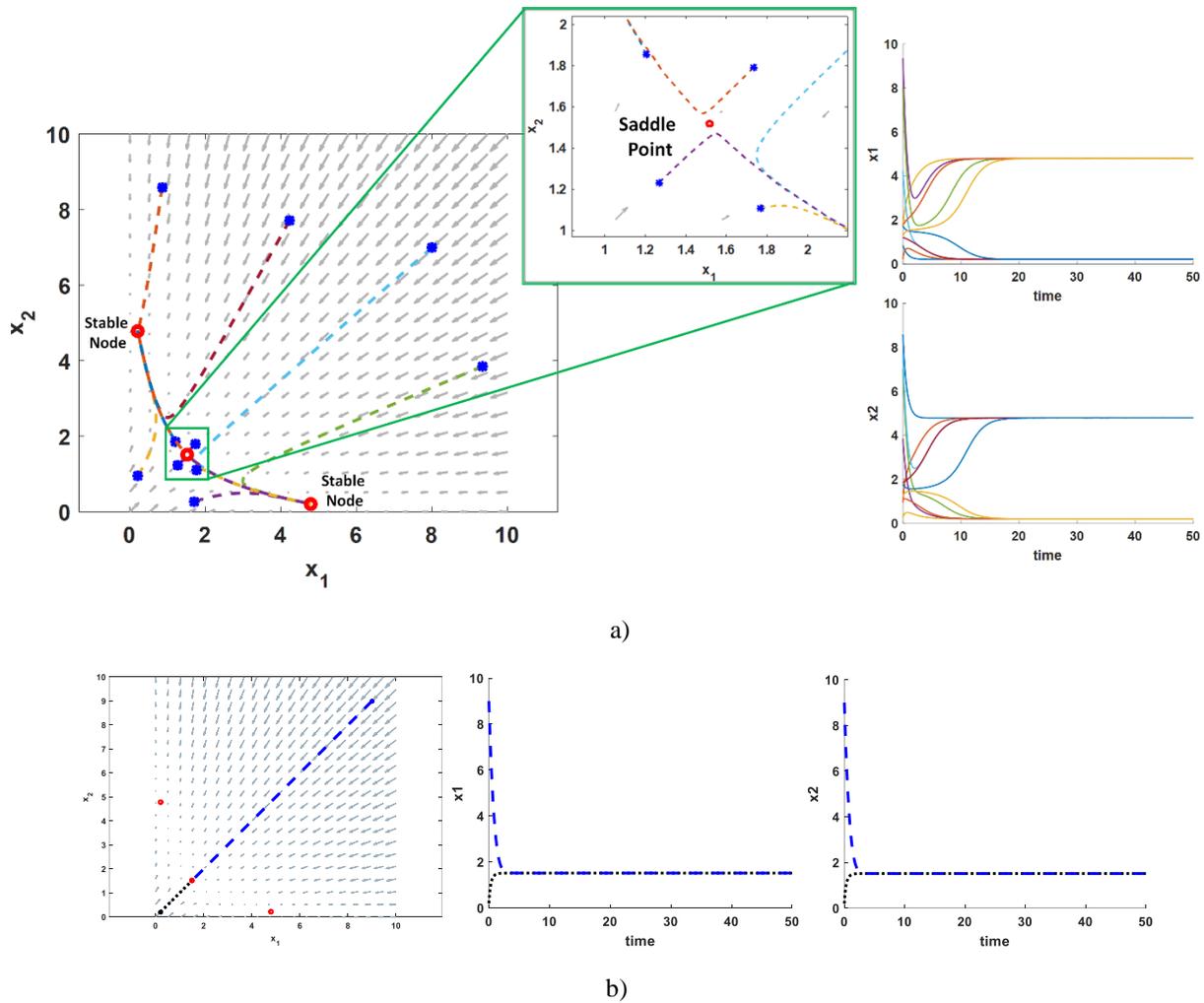

Fig.2. a) The phase portrait, the equilibrium points and the examples of phase trajectories from different initial states in the deterministic model of GTS ($\alpha = 5, \beta = 2, \rho = 1$), b) The stable manifold around the saddle point.

## 2.1. The Impact of Genetic and Epigenetic Alterations

Regarding the huge amount of uncertainty in molecular environment, a real GRN is always faced to two basic kinds of subcellular variations including genetic and epigenetic alterations. Computer simulations confirm that genetic alterations can drastically change the equilibrium states, and even in some cases depending on the mutation intensity, the dimension of the equilibrium state may be altered. Also, epigenetic alterations can change the equilibrium states but more smoothly than genetic alterations. In general, it is exposed that both factors can play a crucial role in the determination of cell fate and accordingly alter the differentiation, development, and phenotypic characteristics.

In case of GTS, the systemic parameters of this circuit is highly dependent to the environmental factors and genetic/epigenetic alterations, and thus, none of two GTSs have same parameters. It is very important to note that the nonlinear dynamics of GTS can be drastically affected by these alterations. For example, Fig.3 reveals that the number of equilibrium points, the dynamics of the phase portrait, and particularly the bistability property of GTS are all altered by changing the circuit parameters. As shown in this figure (10 random initial states), when $\alpha = 1$, the circuit has only one stable equilibrium point at (0.68, 0.68). As another example, assume that the second gene is changed by mutation, then the protein translated from this gene will not have its original property and thus is cannot repress the transcription of the first gene. Computationally, it means that $x_2$ is equal to 0 in Eq.1. As a result, the attractor will be drastically changed.

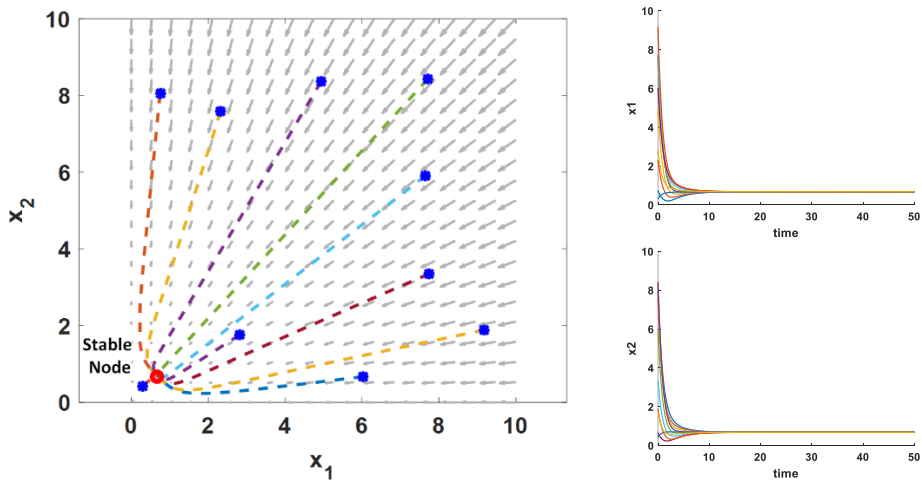

Fig.3. The impact of the parameter values on the bistability property of GTS ($\alpha = 1, \beta = 2, \rho = 1$)

What is a robust computational model for a GTS which can describe the dynamics of the circuit under such uncertain alterations? This paper employs the notion of uncertain stochastic differential equations to answer this question. In the rest of this section, first the effect of each fundamental type of uncertainty, i.e. probabilistic and possibilistic, is considered separately. Then, the proposed model is explained.

### 2.1.1. Probabilistic Uncertainty

Using the notion of stochastic differential equations (SDEs), probabilistic uncertainty can be incorporated into the deterministic model of Eq.1 by adding a noise term $\eta$ as follows:

$$\begin{cases} \frac{dx_1}{dt} = \frac{\alpha}{1+x_2{}^\beta} - \rho x_1 + \eta_1(t) \\ \frac{dx_2}{dt} = \frac{\alpha}{1+x_1{}^\beta} - \rho x_2 + \eta_2(t) \end{cases} \tag{2}$$

Fig.4 shows a comparison between the deterministic response of Eq.1 and the stochastic responses of Eq.2 for 10 independent random runs, but all from a same initial state. The results clearly demonstrate that noise can drastically change the cell fate, such that the responses of Eq.2 are completely different with the prediction of the deterministic model in about 50% of the trials.

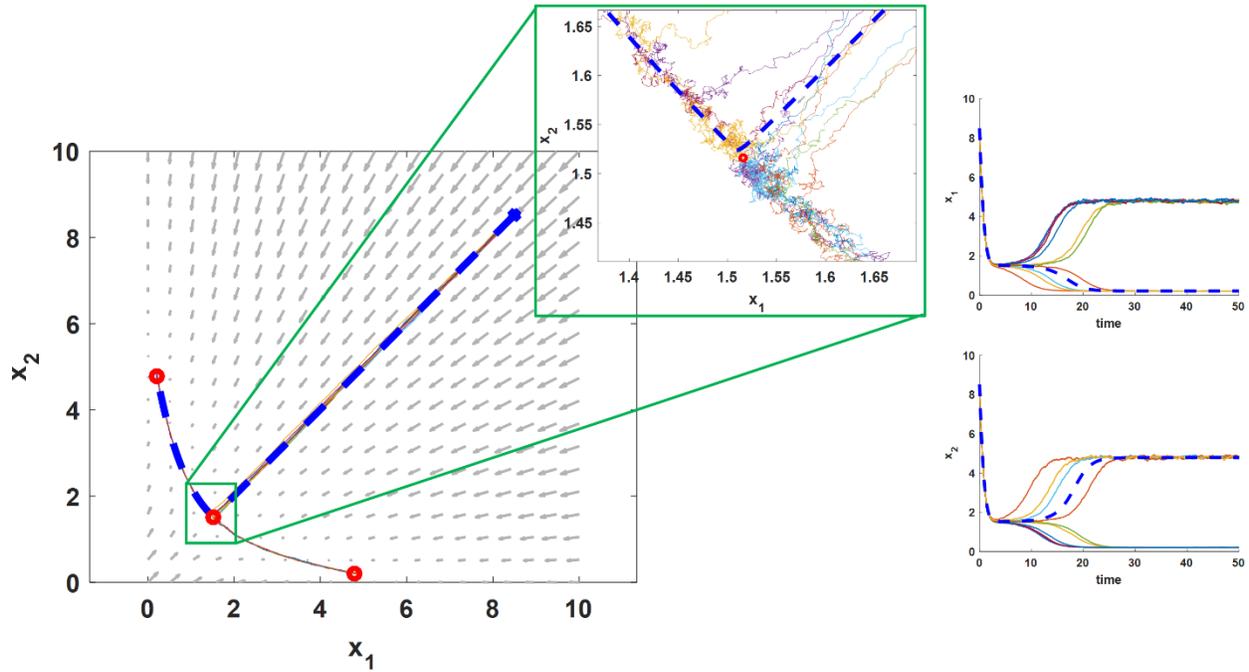

Fig.4. A comparison between deterministic response (dashed) and stochastic responses (solid) from same initial state ($\alpha = 5, \beta = 2, \rho = 1, \eta: 10\%$)

### 2.1.2. Possibilistic Uncertainty

The environmental factors and genetic/epigenetic alterations can highly affect the values of the systemic parameters of GTS. In real-world, none of two similar GTSs (biological replicates) have same parameter values. Consequently, different phenotypes may be observed in the cells of a same type only due to parametric uncertainty. The results of bifurcation analysis of GTS in Fig.5 reveals that the number and the location of the equilibrium points are highly sensitive to the values of parameters.

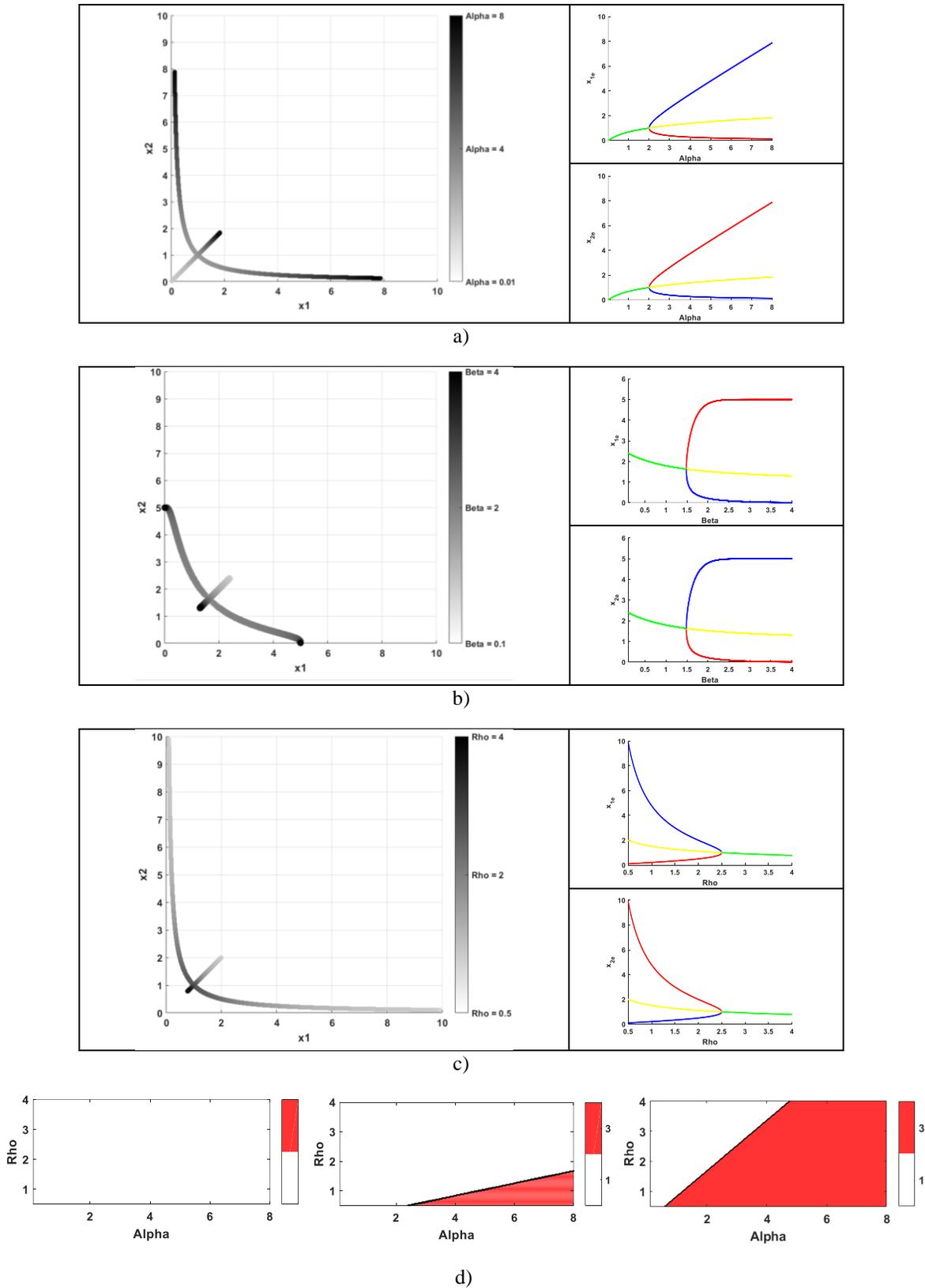

Fig.5. Bifurcation Analysis of GTS: The location of equilibrium points versus a) $\alpha$ where $\beta = 2$ and $\rho = 1$, b) $\beta$ where $\alpha = 5$ and $\rho = 1$, c) $\rho$ where $\alpha = 5$ and $\beta = 2$, d) The relationship between the number of equilibrium points and the values of parameters, where $\beta$ is set to 0.5, 1.5, 3 from left to right.

The parametric uncertainty can be incorporated into the deterministic model of Eq.1 using the notion of Uncertain ODEs as follows:

$$\begin{cases} \frac{dx_1}{dt} = \frac{(\alpha+\Delta\alpha_1)}{1+x_2^{(\beta+\Delta\beta_1)}} - (\rho + \Delta\rho_1)\,x_1 \\ \frac{dx_2}{dt} = \frac{(\alpha+\Delta\alpha_2)}{1+x_1^{(\beta+\Delta\beta_2)}} - (\rho + \Delta\rho_2)\,x_2 \end{cases} \quad (3)$$

The results of computer simulation in Fig.6 depicts that parametric uncertainty can significantly changes the dynamics of GTS and the pattern of phase portrait. In all diagrams, the initial state is same, but the phase trajectories (phenotype) and the final points (fate) are completely different.

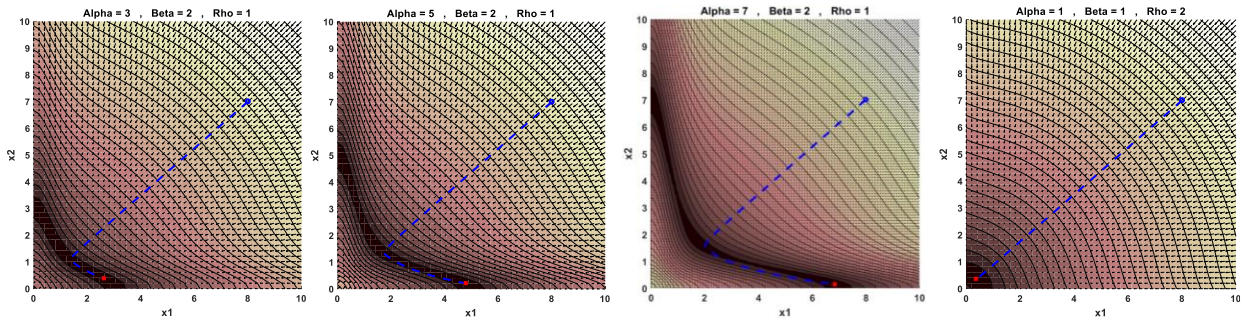

Fig.6. The effect of parametric uncertainty on the phenotype and fate of GRN

## 2.2. The Proposed Model

Since in real GRNs, both types of uncertainty affect the dynamics of the circuit, this paper uses *Uncertain Stochastic Differential Equations* as a robust approach for computational modeling of GRNs. Based on this idea, a robust model of GTS can be described as follows:

$$\begin{cases} \frac{dx_1}{dt} = \frac{\alpha+\Delta\alpha_1}{1+x_2^{\beta+\Delta\beta_1}} - (\rho + \Delta\rho_1)x_1 + \eta_1(x_1.x_2) \\ \frac{dx_2}{dt} = \frac{\alpha+\Delta\alpha_2}{1+x_1^{\beta+\Delta\beta_2}} - (\rho + \Delta\rho_2)x_2 + \eta_2(x_1.x_2) \end{cases} \quad (4)$$

Fig.7 displays the phase trajectories and the temporal gene expression profiles of the GTS of Eq.4 in 4 distinguishing tests with different values of parameters, initial state, and rate of possibilistic/probabilistic uncertainty. In each test, temporal gene expression profiling has been performed from the GTS of Eq.4 through 20 independent runs in in-silico study. Computer simulations reveal that the response of the proposed model is close to the behavior of actual GTSs, in which, real-world uncertainty play critical role in changing the patterns of phase portrait and temporal gene expression profile. As shown in Fig.7, these patterns are very distinguishing and confusing in comparison with the nominal deterministic model. Now, the crucial question is how one can identify the attractors of an actual GTS from such puzzling temporal gene expression profiles in the presence of uncertainty? This paper addresses this question in the next section.

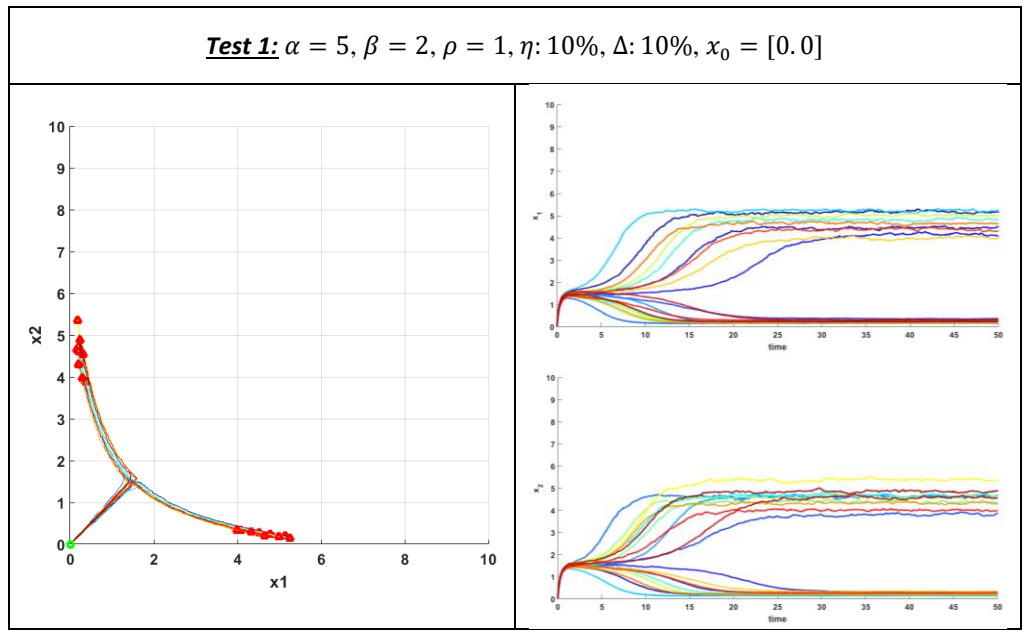

Test 1: $\alpha = 5, \beta = 2, \rho = 1, \eta: 10\%, \Delta: 10\%, x_0 = [0.0]$

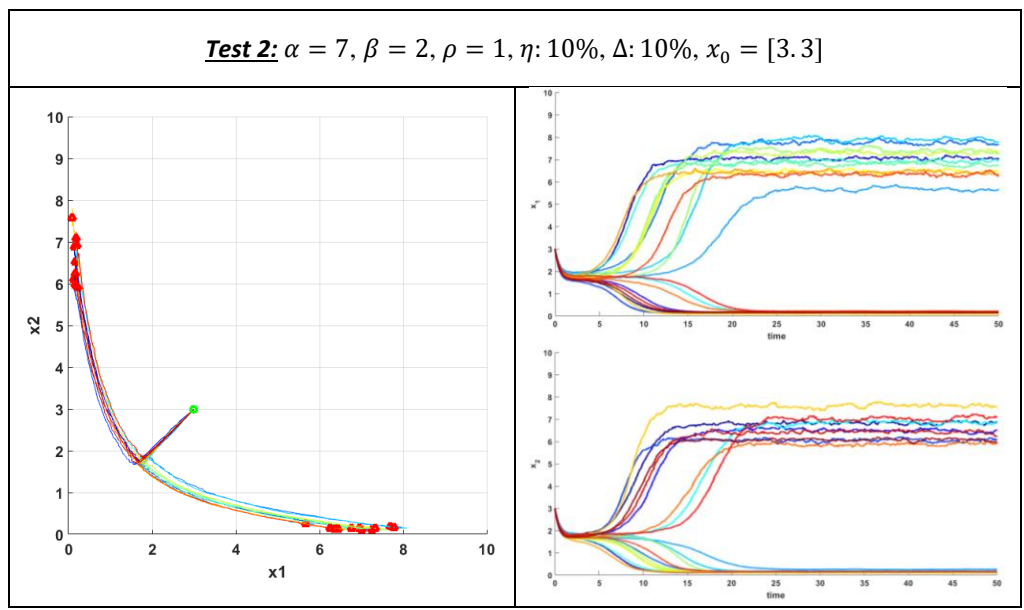

Test 2: $\alpha = 7, \beta = 2, \rho = 1, \eta: 10\%, \Delta: 10\%, x_0 = [3.3]$

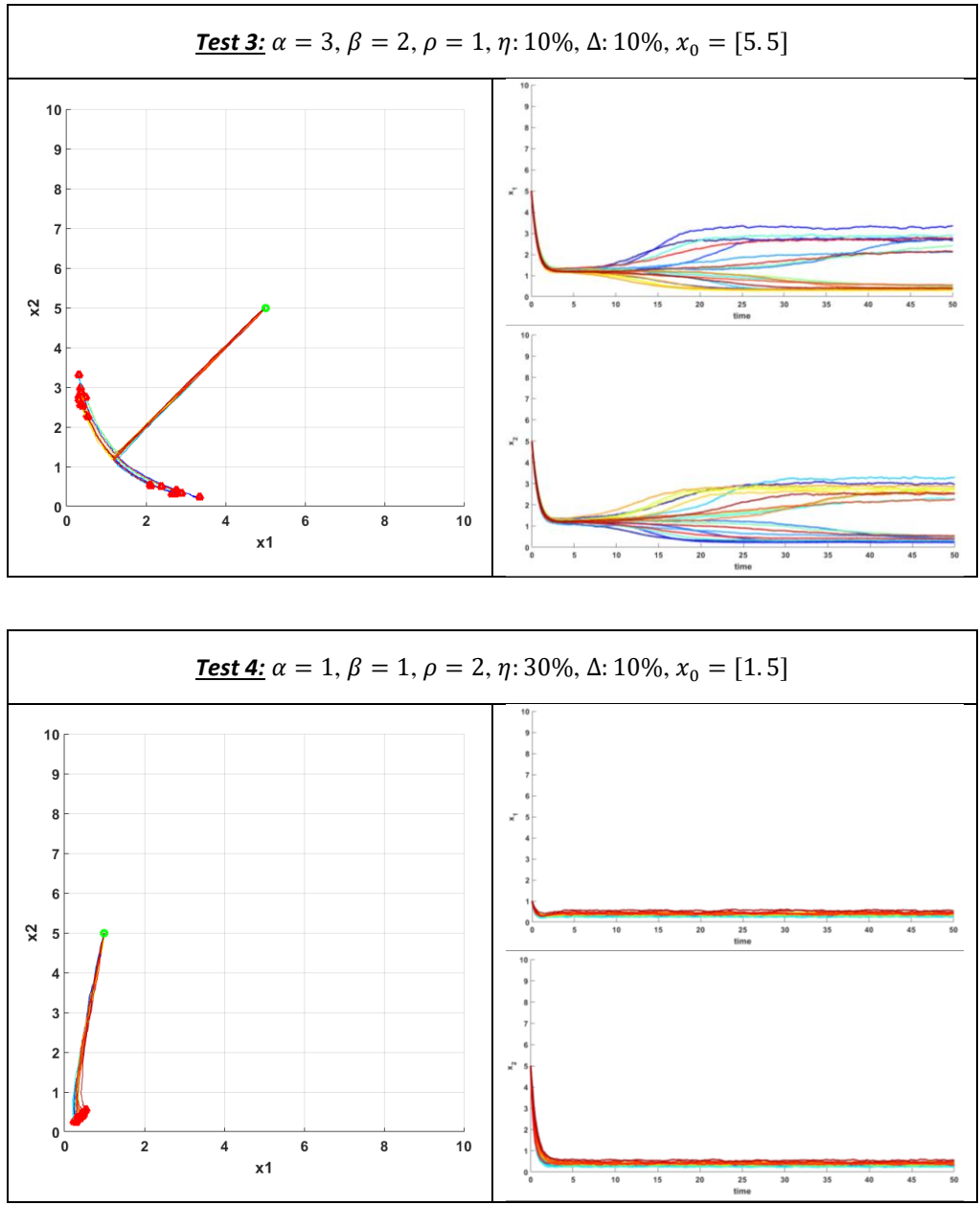

Fig.7. The phase trajectories and temporal gene expression profiles of the GTS of Eq.4 in 4 distinguishing tests.

## 3. Fuzzy Logic-based Attractor Landscape Analysis

In [26], the author proposed a novel approach for the analysis of time-course gene expression data based on Zadeh's computing with words. The proposed scheme could efficiently discover patterns from temporal gene expression profiles in terms of both *fuzzy-logic-based granular* and *linguistic* descriptions which is simply interpretable by human experts. Therefore, this method can be considered as an effective step towards interpretable AI. Also, an open-source software package entitled 'FuzzGene' has been developed for implementation of the proposed method in [26]. In the present paper, a similar approach is employed to identify the attractors of the GTS of Eq.4 from a limited temporal gene expression profiles in the presence of uncertainty. This method is general and can be applied to other biological circuits.

First, using the notion of fuzzy granulation, the state space of the computational model is partitioned into a number of fuzzy granules, where a fuzzy granule is defined as the intersection of the fuzzy sets defined over the state variables. Fig.8 depicts how the state space of Eq.4 is partitioned by fuzzy granules. Then, a fuzzy-logic-based procedure is designed to convert the time series of the temporal gene expression profiles to one fuzzy signal which could efficiently handle the uncertainty available in the profiling process. Finally, a fuzzy scheme is developed to analyze the fuzzy signal and its steady state value is assigned to few fuzzy granules with computing the firing strength. The fuzzy granules with the firing strength of above a predefined threshold can be considered as the fuzzy attractors of the circuit and the firing strength can be introduced as a confidence metric to determine the robustness of the identified attractors.

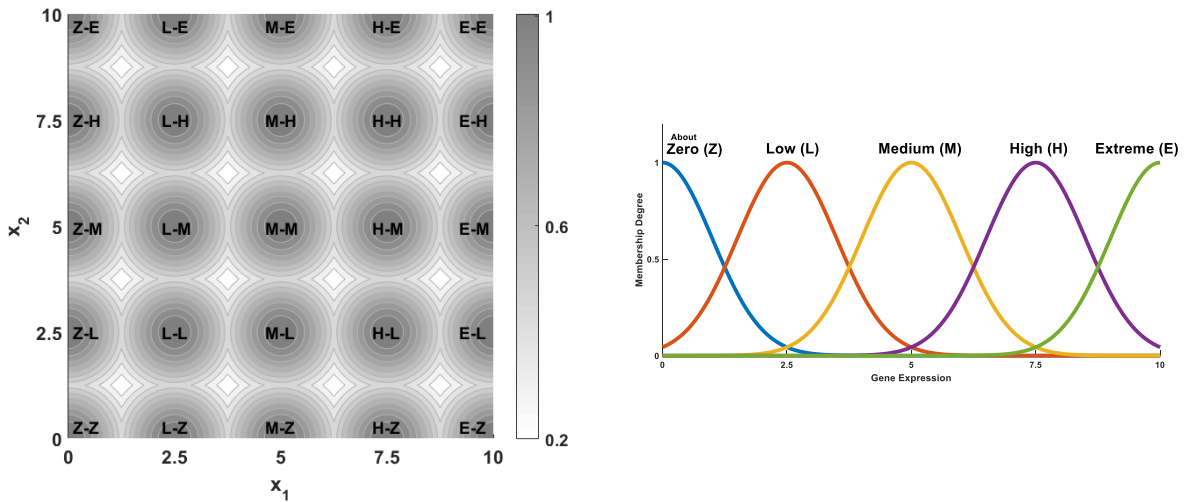

Fig.8. Fuzzy Granulation of State Space: a) Fuzzy granules on the state space of Eq.4, b) Fuzzy sets over each state variable.

Fig.9 shows the phase trajectories of Test 1 to 4 of the previous section on the fuzzy granular state space, where the color map stands for the membership degree of fuzzy granule. In this paper, the threshold is set to be *0.2*. Fig.10 demonstrates that the proposed approach could robustly identify the attractors of GTS in Test 1 to 4 in terms of interpretable linguistic results. Also, the basin of attraction, the type of attractor, and the attraction potential can be approximated similarly by fuzzy logic from the transient values of the fuzzy signals [26-27].

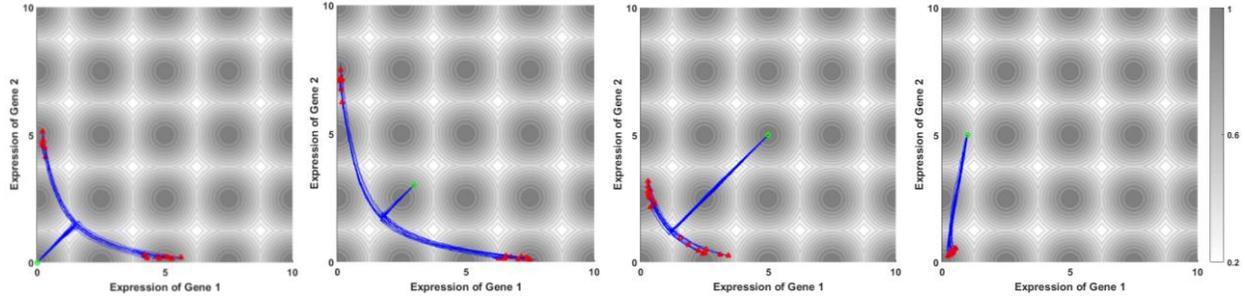

Fig.9. The phase trajectories of Test 1 to 4 (from left to right) in the Fuzzy Granular State Space.

| Test 1 | Transient Response | | | | | | | | | | Steady-state Response | |
|---|---|---|---|---|---|---|---|---|---|---|---|---|
| Time | 0 – 0.35 | 0.35 – 0.38 | 0.38 – 0.55 | 0.55 – 0.61 | 0.61 – 0.93 | 0.93 – 10.38 | 10.38 – 11.68 | 11.68 – 11.81 | 11.81 – 14.33 | 14.33 – 14.62 | 14.62 – 16.54 | 16.54 – 50 | 50 |
| Granules with Confidence | Z-Z 1 | Z-Z 0.35 L-Z 0.20 | Z-Z 0.32 L-Z 0.20 Z-L 0.20 | Z-Z 0.22 L-Z 0.21 Z-L 0.21 | Z-Z 0.21 L-Z 0.21 L-L 0.22 | L-Z 0.21 Z-L 0.21 L-L 0.30 | L-Z 0.30 L-L 0.30 | L-Z 0.27 Z-L 0.20 L-L 0.25 | L-Z 0.27 M-Z 0.20 Z-L 0.22 L-L 0.20 | L-Z 0.27 M-Z 0.21 Z-L 0.22 | M-Z 0.32 Z-L 0.21 | M-Z 0.34 | M-Z 0.43 Z-M 0.20 | M-Z 0.52 Z-M 0.35 |
| Interpretable Result | **Identified Attractors:**<br>1. *Expression of Gene 1* is <u>Medium</u> & *Expression of Gene 2* is <u>About Zero</u><br>2. *Expression of Gene 1* is <u>About Zero</u> & *Expression of Gene 2* is <u>Medium</u> | | | | | | | | | | | | |

| Test 2 | Transient Response | | | | | | | Steady-state Response | |
|---|---|---|---|---|---|---|---|---|---|
| Time | 0 – 6.19 | 6.19 – 8.29 | 8.29 – 10.17 | 10.17 – 10.61 | 10.61 – 13 | 13 – 13.42 | 13.42 – 13.88 | 13.88 – 50 | 50 |
| Granules with Confidence | L-L 0.78 | L-Z 0.20 L-L 0.59 | L-L 0.36 | M-Z 0.20 L-L 0.23 | M-Z 0.21 | M-Z 0.21 Z-H 0.20 | Z-H 0.21 | H-Z 0.20 Z-H 0.23 | H-Z 0.42 Z-H 0.38 |
| Interpretable Result | **Identified Attractors:**<br>1. *Expression of Gene 1* is <u>High</u> & *Expression of Gene 2* is <u>About Zero</u><br>2. *Expression of Gene 1* is <u>About Zero</u> & *Expression of Gene 2* is <u>High</u> | | | | | | | | |

| Test 3 | Transient Response | | | | | | | | | | Steady-state Response | |
|---|---|---|---|---|---|---|---|---|---|---|---|---|
| Time | 0 – 0.24 | 0.24 – 0.25 | 0.25 – 0.29 | 0.29 – 0.31 | 0.31 – 0.36 | 0.36 – 1.62 | 1.62 – 1.68 | 1.68 – 2.32 | 2.32 – 2.94 | 2.94 – 14.05 | 14.05 – 50 | 50 |
| Granules with Confidence | M-M 1 | L-M 0.20 M-M 0.34 | M-L 0.20 L-M 0.20 M-M 0.31 | L-L 0.20 M-L 0.21 L-M 0.21 M-M 0.22 | L-L 0.24 M-L 0.21 L-M 0.21 | L-L 0.35 | L-Z 0.20 L-L 0.34 | L-Z 0.20 Z-L 0.20 L-L 0.20 | Z-Z 0.20 L-Z 0.21 Z-L 0.21 L-L 0.22 | Z-Z 0.22 L-Z 0.21 Z-L 0.21 | L-Z 0.22 Z-L 0.27 | L-Z 0.31 Z-L 0.54 |
| Interpretable Result | **Identified Attractors:** <br> 1. *Expression of Gene 1 is Low  &  Expression of Gene 2 is About Zero* <br> 2. *Expression of Gene 1 is About Zero  &  Expression of Gene 2 is Low* | | | | | | | | | | | |

| Test 4 | Transient Response | | | | | Steady-state Response | |
|---|---|---|---|---|---|---|---|
| Time | 0 – 0.09 | 0.09 – 0.11 | 0.11 – 0.23 | 0.23 – 0.58 | 0.58 – 1.41 | 1.41 – 50 | 50 |
| Granules with Confidence | Z-M 0.61 L-M 0.32 | Z-M 0.50 | Z-L 0.22 Z-M 0.45 | Z-L 0.60 | Z-Z 0.21 Z-L 0.69 | Z-Z 0.74 | Z-Z 0.87 |
| Interpretable Result | **Identified Attractor:** <br> *Expression of Gene 1 is About Zero  &  Expression of Gene 2 is About Zero* | | | | | | |

Fig.10. Robust identification of the attractors of GTS in terms of interpretable linguistic results

## 4. Conclusions

This paper proposed a novel approach based on Zadeh's *Computing with Words* for robust identification of the attractors of GRNs from temporal gene expression data in the presence of uncertainty. In addition to fuzzy logic-based computational results, the proposed method could generate linguistic results which is simply interpretable by human experts. Therefore, this scheme can be considered as an effective step towards interpretable AI. As a case study, the benchmark GRN of GTS was considered, which is able to perform logical functions at molecular level and it has been recognized as a building block for construction of more complex GRNs. However, this method is general and can be applied to other biological circuits. The impact of genetic/epigenetic alterations as well as varying environmental factors on changing the location, characteristics and even the number of attractors were discussed. In order to handle both types of probabilistic and possibilistic uncertainty, the nonlinear dynamics of GTS was modeled based on the notion of uncertain stochastic differential equations. Bifurcation analysis and the sensitivity of phase portrait and bi-stability property to uncertainty were studied by computer simulation. The results of distinguishing tests through in-silico study demonstrated that the proposed fuzzy logic-based approach could robustly identify the attractors of GTS from temporal gene expression data in the presence of uncertainty and generated linguistic results interpretable for human inference.